\documentclass[twocolumn,preprintnumbers, amssymb,amsmath,aps,floatfix,prl,nofootinbib,superscriptaddress,showpacs]{revtex4-1}
\usepackage{epsfig}
\usepackage{bm}
\usepackage{amssymb}
\usepackage{amsmath}
\usepackage{soul,xcolor}
\usepackage{color}
\usepackage[colorlinks,
            linkcolor=blue,
            anchorcolor=black,
            citecolor=blue
            ]{hyperref}
\newcommand{\beq}{\begin{eqnarray}}
\newcommand{\eeq}{\end{eqnarray}}
\setstcolor{red}

\newcommand{\Eq}[1]{Eq.(\ref{#1})}

\newcommand{\Fig}[1]{Fig.\,\ref{#1}}

\begin{document}
\title{QCD Resummation in Hard Diffractive Dijet Production at the Electron-Ion Collider}

\author{Yoshitaka Hatta}
\affiliation{Physics Department, Building 510A, Brookhaven National Laboratory, Upton, NY 11973}

\author{Niklas Mueller}
\affiliation{Physics Department, Building 510A, Brookhaven National Laboratory, Upton, NY 11973}

\author{Takahiro Ueda}
\affiliation{Department of Materials and Life Science, Seikei University
3-3-1 Kichijoji Kitamachi, Musashino-shi, Tokyo 180-8633, Japan}
\author{Feng Yuan}
\affiliation{Nuclear Science Division, Lawrence Berkeley National
Laboratory, Berkeley, CA 94720, USA}

\begin{abstract}
%\nm{One introductory sentence here.}
Diffractive dijet production at the electron-ion collider (EIC) has been proposed to study the gluon Wigner distribution at small-$x$. We investigate the soft gluon radiation associated with the final state jets and an all order resummation formula is derived. We show that the soft gluon resummation plays an important role to describe E791 data on $\pi$-induced diffractive dijet production at Fermilab. Predictions for the EIC are presented, and we emphasize that the soft gluon resummation is an important aspect to explore the nucleon/nucleus tomography through these processes. 
\end{abstract}
%\pacs{24.85.+p, 12.38.Bx, 12.39.St, 12.38.Cy}
\maketitle

{\it Introduction.}
There have been renewed interests in hard diffractive dijet production in $e+p$ and $e+A$ collisions, which was one of the focuses of previous theoretical studies decades ago~\cite{Nikolaev:1994cd,Bartels:1996ne, Bartels:1996tc, Diehl:1996st, Bartels:1999tn, Braun:2005rg,Rezaeian:2012ji, Marquet:2007nf, GolecBiernat:2005fe}. It was triggered by the possibility to explore the parton Wigner distributions in these processes~\cite{Altinoluk:2015dpi,{Hatta:2016dxp},{Ji:2016jgn},{Hatta:2016aoc},Hagiwara:2017fye,Mantysaari:2019csc,Salazar:2019ncp}. The Wigner distributions of quarks and gluons~\cite{Ji:2003ak, Belitsky:2003nz} represent an important aspect of the tomographic study for nucleons and nuclei in recent years, which is also one of the major focuses at the planned electron-ion colliders (EIC)~\cite{Boer:2011fh, AbelleiraFernandez:2012cc, Accardi:2012qut}. 

One of the key observations in the new proposal is to measure the total transverse momentum of the dijet, the Fourier transform of which provides information on the coordinate space distribution of the partons. Together with the individual jet transverse momentum, this leads to a multi-dimensional tomographic picture of the nucleons and nuclei. %Since the total transverse momentum of the dijet depends on the diffractive slope, it is limited to small transverse momentum region. \nm{unclear what this means. references?} 
Therefore, a precise measurement of the total transverse momentum distribution is of crucial importance to measure quark and gluon Wigner distribution functions.

Most previous analyses were based on the leading order picture of diffractive dijet production. To consolidate the factorization property of this process, we need to investigate higher order perturbative corrections~\cite{Boussarie:2016ogo,Boussarie:2019ero} and the relevant QCD evolution effects~\cite{Echevarria:2016mrc,Mantysaari:2019csc}. In this paper, we will consider one of the important higher order contributions, i.e., all order soft gluon radiation associated with the final state jets. They can strongly affect the transverse momentum distribution of the dijet system at low momentum, and this should be taken into account when extracting the coordinate space distribution.   %Therefore, we need to understand the soft gluon radiation to fully explore the hard diffractive dijet production.

In order to change the dijet transverse momentum, the soft gluons have to be emitted outside the jet cones. They are therefore insensitive to the collinear singularity, and the relevant resummation becomes single logarithmic. It is known that such a resummation consists of two parts---the Sudakov logarithms and the so-called non-global logarithms (NGLs)~\cite{Dasgupta:2001sh,Dasgupta:2002bw,Banfi:2002hw,Banfi:2003jj,Hatta:2009nd,Hatta:2013iba,Neill:2018mmj}. 
%The comparison of the Sudakov resummation effects between the non-diffractive and diffractive processes will provide a unique opportunity to study the associated QCD dynamics. 
The resummation of Sudakov logarithms is straightforward, and it will be interesting to compare their impacts in diffractive and non-diffractive processes. 
The resummation of NGLs, on the other hand, is known to be quite nontrivial, but to leading logarithmic approximation it can be done by using the existing techniques. 
%Among these contributions, both Sudakov-type logarithms and the so-called non-global logarithms (NGLs) play important roles in the description of jet related observables.

%\nm{Maybe, one short paragraph here, were the soft radiation has been studied in other context. references. comparison with experiment}

In our study, we will only consider  color-neutral particles in the initial state (i.e., not a single quark or a gluon from the incoming hadrons) and color-singlet $t$-channel exchanges. At the EIC, the incoming electron radiates a virtual photon which diffractively scatters off the nucleon target and produces two final state jets. Another example are pion-induced coherent diffractive dijet processes, studied in fixed target experiments~\cite{Frankfurt:1993it,{Aitala:2000hb},{Aitala:2000hc},Nikolaev:2000sh,Frankfurt:1999tq,Frankfurt:2000jm,Braun:2001ih,Braun:2002wu,Chernyak:2001ph,Chernyak:2001wk}. These two examples share strong similarities in the soft gluon radiation contributions and we will compare our resummation formula to existing data from the E791 experiment~\cite{{Aitala:2000hb},{Aitala:2000hc}}. This will provide a benchmark test to assess the applicability of our approach to  diffractive dijet production at the EIC. 

The rest of this paper is organized as follows. First, we will derive the soft gluon resummation contribution to the diffractive dijet production processes. We include both Sudakov and NGL contributions based on the Banfi-Marchesini-Smye (BMS) evolution equation~\cite{Banfi:2002hw}. Because of the universality of the soft gluon resummation, we apply our formula to the $\pi$-induced diffractive dijet processes. We will show that resummation plays an important role in the description of experimental data. We then apply our formalism for predictions at the EIC, where we will show the resummation effects on the momentum distribution and the azimuthal angular asymmetry. The latter is of particular interesting because it provides a novel correlation in the small-$x$ gluon Wigner distribution. Finally, we summarize our paper. 

%\section{Soft Gluon Resummation in Diffractive Dijet Processes}
{\it Soft Gluon Resummation in Diffractive Dijet Processes.}
We start with the leading order cross section of diffractive dijet production in $ep$ and $eA$ collisions 
%. has been formulated before, and can be summarized as follows,
\begin{equation}\label{eq:leadingorder}
    \frac{d\sigma}{d\Omega }=\int d^2\Delta_\perp \frac{d\sigma_0(y_1,y_2;P_\perp,\Delta_\perp)}{dy_1dy_2d^2P_\perp d^2\Delta_\perp} \delta^{(2)}(q_\perp+\Delta_\perp)\ ,
\end{equation}
where $d\Omega=dy_1dy_2 d^2k_{1\perp}d^2k_{2\perp}$ represents the phase space for the two final jets with  rapidities $y_{1,2}$ and transverse momenta $k_{1\perp}$ and $k_{2\perp}$, respectively. $\vec{P}_\perp= (\vec{k}_{1\perp}-\vec{k}_{2\perp})/2$ is the relative transverse momentum of the two jets and the total transverse momentum is defined as $\vec{q}_\perp=\vec{k}_{1\perp}+\vec{k}_{2\perp}$. In the leading order kinematics, $\vec{q}_\perp=-\vec{\Delta}_\perp$ where $\Delta_\perp$ is the transverse component of the nucleon recoil momentum.  In the so-called correlation limit, $\vec{k}_{1\perp}\approx -\vec{k}_{2\perp}$, and we choose $P_\perp\sim |\vec{k}_{1\perp}|\sim  |\vec{k}_{2\perp}|\gg |\vec{q}_\perp|$ to represent the jet transverse momentum. 
%The total transverse momentum $q_\perp$ is much smaller than $P_\perp$. 
In the forward kinematics $y_{1,2}\gg 1$ which corresponds to the small-$x$ region of the nucleon/nucleus, the cross section $\sigma_0$ can be written as the convolution of the hard kernel and the gluon Wigner distributions~\cite{{Hatta:2016dxp},Mantysaari:2019csc}. 

In experiments, if the final state nucleon momentum $P'$ can be re-constructed, one can directly measure the $\Delta_\perp$-distribution which provides information on parton distributions in  impact parameter space. Alternatively, and complementarily, if one tries to reconstruct the $\Delta_\perp$-dependence from the measurement of the $q_\perp$-distribution of the two jets, one has to take into account the additional soft gluon radiation contribution. %This is exactly what the E791 collaboration has done to identify the momentum transfer in the nucleus targets

\begin{figure}
\includegraphics[width=8cm]{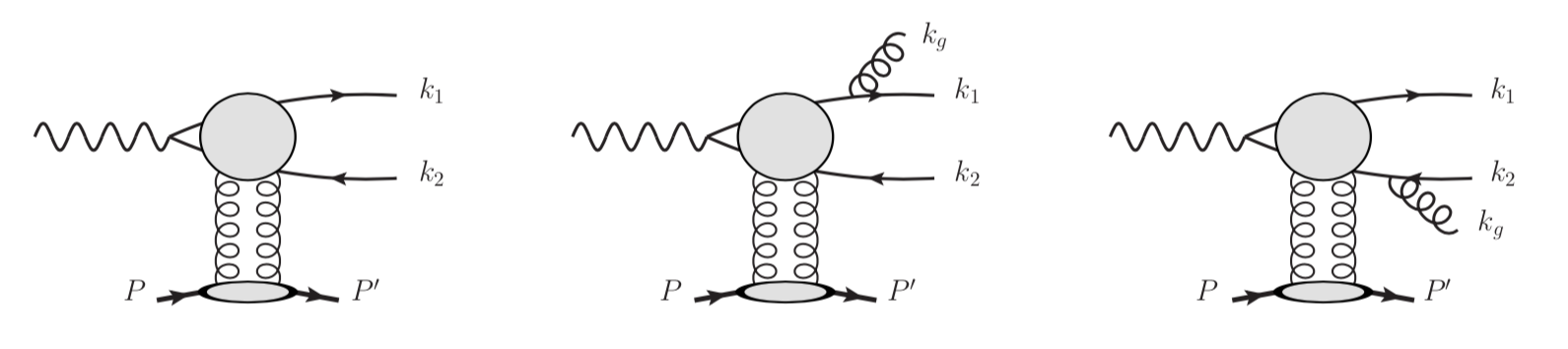}
\caption{Schematic diagrams for hard diffractive dijet production in $e+p$ and $e+A$ collisions. Soft gluon radiation associated with the jets in the final state will contribute to  the leading power  at small total transverse momentum of the dijet.}
\label{dff}
\end{figure}

Due to the colorless exchange in the $t$-channel, the soft gluon radiation associated with the final state jets is very similar to that in jet production in $e^+e^-$ annihilation. Typical one-gluon radiation diagrams are shown in Fig.~\ref{dff}.  Since only the soft radiations emitted outside the jet cones count, the relevant resummation is single-logarithmic, of the sort studied in Ref.~\cite{Dasgupta:2001sh,Dasgupta:2002bw} where large logarithms come from both the Sudakov and non-global effects. 
To leading logarithmic accuracy in the large-$N_c$ approximation, the resummation of both these logarithms can be done by performing Monte-Carlo simulations \cite{Dasgupta:2001sh,Dasgupta:2002bw}, or solving a differential equation called the BMS equation~\cite{Banfi:2002hw}. At finite $N_c$, this can be done by the  Langevin simulation of SU($N_c$) matrices \cite{Hatta:2013iba}. 
% In particular, the NGLs is not easy to resum without BMS equation.

Taking into account the soft gluon radiation contributions, we can re-write the differential cross section as
\begin{eqnarray}\label{eq:convolution}
    \frac{d\sigma}{d\Omega}&=&\int d^2\Delta_\perp \frac{d\sigma_0(y_1,y_2;P_\perp,\Delta_\perp)}{dy_1dy_2d^2P_\perp d^2\Delta_\perp}\frac{S(|q_\perp+\Delta_\perp|)}{2\pi |q_\perp + \Delta_\perp|} \nonumber \\
    &=&\int d^2\lambda_\perp \frac{d\sigma_0(y_1,y_2;P_\perp,\Delta_\perp)}{dy_1dy_2d^2P_\perp d^2\Delta_\perp}\frac{S(|\lambda_\perp|)}{2\pi |\lambda_\perp|},
    \label{base}
\end{eqnarray}
where the soft factor $S(|\lambda_\perp|)$ represents the probability, normalized as $\int_0^{P_\perp}d\lambda_\perp S(\lambda_\perp)=1$, that the transverse momentum emitted outside the jet cones is exactly $|\lambda_\perp|=|-q_\perp-\Delta_\perp|$. This can be calculated from  $P(\tau)$, the probability that the transverse momentum emitted outside the jet cones is {\it less} than $\lambda_\perp$ where 
\beq \label{eq:taudef}
\tau = \frac{N_c}{\pi} \int_{\lambda_\perp}^{P_\perp} \frac{d\lambda'_\perp}{\lambda'_\perp} \alpha_s(\lambda'_\perp).
\eeq
$S(\lambda_\perp)$ is related to $P(\tau)$ via simple differentiation 
\beq\label{eq:softfactordef}
S(\lambda_\perp) = \frac{dP(\tau)}{d\lambda_\perp} = -\frac{N_c}{\pi\lambda_\perp} \alpha_s(\lambda_\perp) \frac{dP(\tau)}{d\tau}. \label{dif}
\eeq
Let us give  simple analytical estimates of $P(\tau)$ and $S(\lambda_\perp)$. As long as $\tau$ is not too large, which is usually the case in practical applications, $P(\tau)$ is dominated by the Sudakov effects. Suppose that the two jets have the same rapidity $y_1=y_2$ and are exactly back-to-back in azimuth $\vec{k}_{1\perp}=-\vec{k}_{2\perp}$. Due to Lorentz invariance, one can boost this system to the center-of-mass frame of the dijet. Then, up to small corrections which stem from the difference between rapidity and angle variables around midrapidity, one finds
\begin{equation}
    P(\tau)\approx \exp\left(-\tau\ln\frac{1+\cos R}{1-\cos R}\right) \ , \label{found}
\end{equation}
    where $R$ is the jet radius. 
From this we immediately obtain 
\begin{equation}
    S(\lambda_\perp)=\frac{\beta}{\pi\lambda_\perp^2} \left(\frac{\lambda_\perp^2}{P_\perp^2}\right)^{\beta} \ ,\label{e6}
\end{equation}
where $\beta=\frac{\alpha_sN_c}{2\pi}\ln\frac{1+\cos R}{1-\cos R}$. If we expand the above result in $\alpha_s$, we find the following leading order result,
\begin{equation}
S^{(1)}(\lambda_\perp)=\frac{\alpha_sN_c}{2\pi^2}\frac{1}{\lambda_\perp^2}\ln\frac{4}{R^2} \ ,\label{softg}
\end{equation}
in the small-$R$ limit. This of course agrees with a direct calculation of the diagrams shown in Fig.~\ref{dff} after identifying $C_F\approx N_c/2$ in the large-$N_c$ limit. 

Going beyond, we have calculated $P(\tau)$ by numerically solving the BMS equation. The jets are placed back-to-back in azimuth $\phi_2=\phi_1+\pi$, and each jet is delineated by a circle of radius $R$ in the $(y,\phi)$ plane. $P(\tau)$ then depends on $R$ and the difference $|y_1-y_2|$. In order to facilitate the differentiation (\ref{dif}),  we have fitted the result by the same analytical formula used in \cite{Dasgupta:2001sh}.    The result for $R=0.4$ and $y_1=y_2$ is, 
\begin{equation}
P(\tau)=\exp\left(-c_1\tau -c_2\tau^2 \frac{1+(a\tau/2)^2}{1+(b\tau/2)^c}\right),
\end{equation}
with $c_1=3.22$,  $c_2=1.01$,  
$a=0.463$, $ b=0.459$, $c=0.574$. As expected, $c_1$ is rather close to the value $\ln \frac{1+\cos 0.4}{1-\cos 0.4} \approx 3.19$ found in  \Eq{found}. The $c_2$ term is due to the nonglobal logarithms.  
% In the next two sections, we shall present two applications of \Eq{base}. One is a fit of the  experimental result from the E791 collaboration at Fermilab, and the other is predictions for the diffractive dijet production at EIC. 

%\section{Test of the resummation Formula in Pion Induced Diffraction}
{\it Test of the resummation Formula in Pion Induced Diffraction.}
In early 2000s, %\st{there has been a beautiful experiment (E791) of diffractive dijet production in $\pi$-induced scattering, where they use 500GeV $\pi$ beam on the nuclear target (Platiumn and Carbon) at Fermilab} 
%\rep{
the E791 experiment at Fermilab measured diffractive dijet production in pion-induced scattering%, using a 500 GeV pion beam on nuclear Platinum and Carbon targets
~\cite{Aitala:2000hb,Aitala:2000hc}. Its main purpose was to explore the color transparency phenomena of the nuclear target (Platinum and Carbon) and novel parton distribution amplitude in pions~\cite{Frankfurt:1993it}.  To do that, the experiment also measured the total transverse momentum $q_\perp$ of the two jets to select the diffractive events. %\st{This is exactly the same as what we studied above. These experimental data, although decades old, provide a unique opportunity to test our understanding of soft gluon radiation contributions in the diffractive dijet processes.} \rep{
Therefore, the E791 experiment provides a unique opportunity to test our understanding of soft gluon radiation in diffractive dijet production processes. %In the following, we will compare the formulas derived above to the experimental data from E791.

Because of the nuclear targets, the diffractive events in E791 experiment contain both coherent and incoherent contributions. The former involves the whole nucleus and the latter involves nucleons in the nucleus. Accordingly, we can write the differential cross section as, approximately~\cite{Frankfurt:2000jm},
\begin{equation}
    \frac{d\sigma_0^A}{d^2\Delta_\perp}\propto \left[A^2e^{-\frac{R_A^2}{3}\Delta_\perp^2}+ Ae^{-\frac{R_p^2}{3}\Delta_\perp^2}\right]\ ,\label{e9}
\end{equation}
where $A$ is the nuclear number, and the first and second terms represent the coherent and incoherent diffractive contributions, respectively. In the above equation, $R_A\sim A^{1/3} R_p$ and $R_p$ are nuclear and nucleon radii. %For simplicity, we have \st{neglected} \rep{have not explicitly} written the dependence on \st{all} other kinematic variables. 

In order to compute the $q_\perp$-distribution from~\Eq{eq:convolution}, we convolute the $\Delta_\perp$ distribution of \Eq{e9} with the soft factor of \Eq{dif}. In the E791 experiment, the jet transverse momentum is about 2 GeV and a special jet algorithm has been applied without an explicit jet size. Therefore, we decide to present our estimate by assuming $\beta\approx 0.6$ in \Eq{e6}, instead of an exact evaluation of $\beta$ which will depend on the jet size. This choice corresponds to a fixed coupling  $\alpha_s=0.3$ and $R\approx 0.25$. We can perform the convolution of \Eq{eq:convolution} numerically, and find that the following analytic approximation for the final $q_\perp$ distribution describes the data well,
\begin{eqnarray}
    \frac{dN}{d^2q_\perp}&=&{\cal N}P_\perp^{-2\beta}\left[A^2 \left(\frac{3}{R_A^2}\right)^{\beta}{}_1F_1\left(1-\beta,1,-\frac{q_\perp^2R_A^2}{3}\right)\right.\nonumber\\
&&    \left.+A \left(\frac{3}{R_p^2}\right)^{\beta}{}_1F_1\left(1-\beta,1,-\frac{q_\perp^2R_p^2}{3}\right)\right] \,. \label{e8}
\end{eqnarray}
The normalization factor ${\cal N}$ depends on all other kinematic variables and ${}_1F_1$ is the Hypergeometric function. %It is worthwhile to mention that the prefactor $P_\perp^{-2\beta}$, which is an immediate consequence of the convolution with the soft factor \Eq{found}, has profound impact on the final $P_\perp$ distribution as well (see discussions below). 

\begin{figure}
\includegraphics[width=0.45\textwidth]{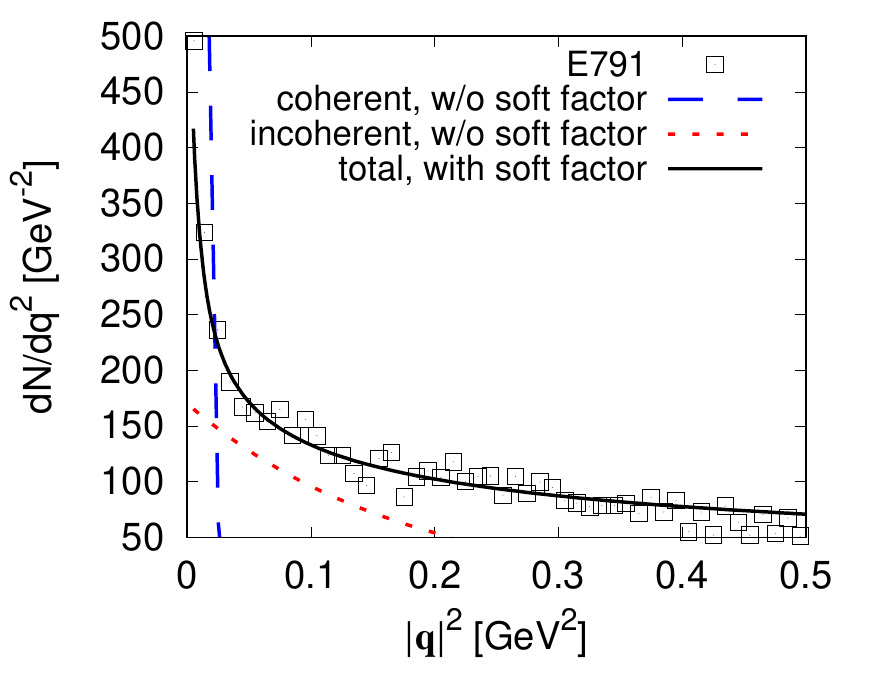}
\includegraphics[width=0.45\textwidth]{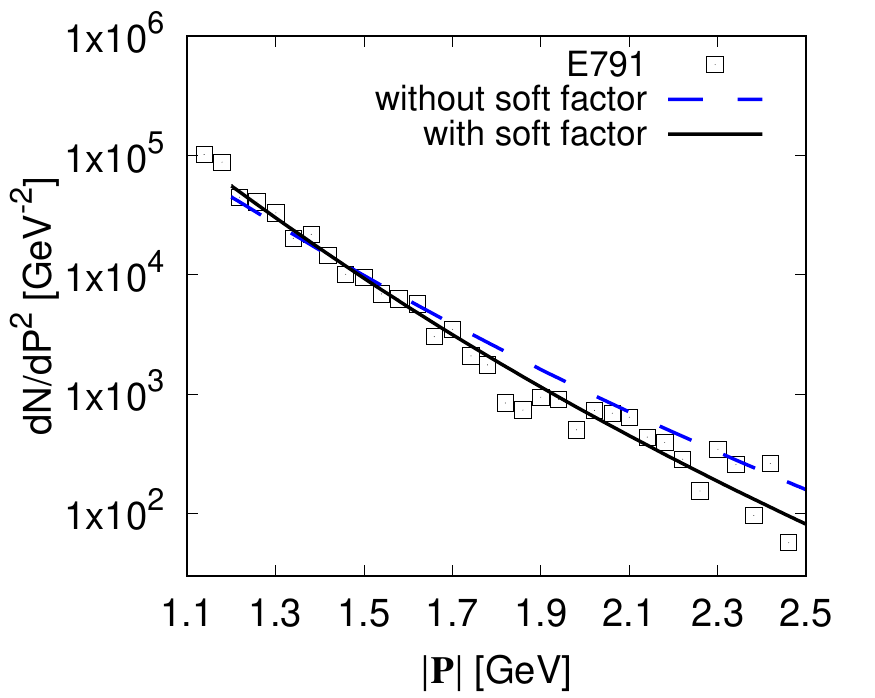}
\caption{Comparisons of the Sudakov effects in the total transverse momentum $q_\perp$ distribution (upper) and jet transverse momentum $P_\perp$ distribution (lower) for the diffractive dijet production in $\pi$-induced scattering on the nuclear target of Platinum with the experimental data from E791 Collaboration~\cite{Aitala:2000hb,Aitala:2000hc}. The normalizations are arbitrary in the comparisons. In the upper plot, the dotted and dashed curves represent the coherent and incoherent diffractive contributions without soft factor, whereas the solid curve is the total contribution with soft factor. In the lower plot, the dashed curve represents the contribution without soft factor and the solid curve with soft factor.}\label{fig:E791}
\label{E791}
\end{figure}

In Fig.~\ref{E791}, we compare our results to the experimental data from E791. %For the numeric calculations, we take $\beta_0=0.6$ in Eq.~(\ref{su0}) where $\beta_0$ is defined as $\beta_0=\alpha_sC_F/\pi \ln(1/R^2)$. We have also taken average jet transvese momentum of 2GeV in the calculations. Because of uncertainties in the experimental determination of the jet radius in these experiments, it is a reasonable assumption to apply a numeric estimate for $\beta_0$. In future experiments, we shall determine $\beta_0$ by directly calculating from the above formula from the jet size for the anti-$k_t$ algorithm. We expect that including running coupling effects shall improve the agreement between the theory and experiment. 
The coherent diffraction dominates at very low $q_\perp$, while the incoherent diffraction starts to take over at moderate $q_\perp$. On the other hand, at relative large $q_\perp$, the soft factor contribution dominates. We emphasize that the soft factor contribution is important for the whole kinematic region of $q_\perp$. Without it, we would not be able to describe the distributions, even at very small-$q_\perp$. We also compared our predictions to experimental data for the Carbon target, and found  agreement using the same $\beta$ parameter. %\footnote{Comparing the two plots in Fig.~2 of Ref.~\cite{Aitala:2000hc}, we find that the relative weight of coherent and incoherent contributions is arbitrary in their fit, which is different from our results that they are fixed by $A^2$ and $A$.}. 
This indicates that the soft gluon radiation is the same as it should be, because it only concerns the jets in the final state.  

We now turn to the jet transverse momentum dependence of the coherent diffractive events, where we show the comparison in the lower plot of Fig.~\ref{E791}. The experimental data are obtained by integrating over $q_\perp^2$ up to $0.015~ {\rm GeV}^2$~\cite{Aitala:2000hb,Aitala:2000hc}. Because the $P_\perp$- and $q_\perp$-dependence are separated in Eq.~(\ref{e8}), the $q_\perp$-integral will not affect the $P_\perp$-dependence. However, the additional factor of $P_\perp^{-2\beta}$ of (\ref{e8}) will enter into final result. From the power counting analysis, the partonic differential cross section leads to a power behavior of $d\hat \sigma/dP_\perp^2 \sim 1/P_\perp^8$~\cite{Frankfurt:2000jm,Braun:2001ih,Braun:2002wu}. By adding additional $P_\perp$-dependence in the associated gluon distribution functions~\cite{Frankfurt:2000jm} and the $1/P_\perp^{2\beta}$ from Eq.~(\ref{e8}), we obtain the theoretical prediction as the solid curve in Fig.~\ref{E791}. The dotted curve are the predictions without soft factor contribution. As shown in \Fig{E791}, the predictions with soft factor have better agreement with the data.

%\section{Predictions for the Electron-Ion-Collider}
{\it Predictions for the Electron-Ion-Collider. }
The comparison between our theory predictions with previous E791 experiment demonstrates the importance of the soft factor contributions. %They play a significant role, in particular, for the total transverse momentum distribution for the final state two jets.
In the following, we will present numeric results for cross sections measurable at a future EIC, focusing on coherent diffractive dijet production in $e+p$ collisions~\cite{Mantysaari:2019csc,Hatta:2016dxp,Altinoluk:2015dpi}.

The cross section can be parametrized by azimuthal Fourier decomposition~\cite{Mantysaari:2019csc}
\begin{align}\label{eq:csazimuthal}
\text{d}\sigma_{L/T} = v_0\big[1+2v_2\cos 2\theta({P}_\perp,{\Delta}_\perp)+\dots\big]
\end{align}
where $\theta({P}_\perp,{q}_\perp)$ is the relative angle between dijet momentum ${P}_\perp = ({k}_{1,\perp}-{k}_{2,\perp})/2$ and the nucleon recoil ${\Delta}_\perp $,
and $L$($T$) denotes a virtual photon with longitudinal (transverse) polarization. A non-zero $v_2$ in diffractive dijet production signals a non-trivial correlation between impact parameter and transverse momentum of the gluon Wigner distribution at small $x$
and is an important benchmark measurement for the EIC~\cite{Mantysaari:2019csc}.

As discussed before, soft gluon radiation of the dijets which is not captured by the jet reconstruction, is an important issue to reconstruct  \Eq{eq:csazimuthal}. In the following, we consider all-order re-summation of soft-gluon radiation and investigate its effects both on the magnitude of the dijet production cross section as well as its elliptic azimuthal modulation. %This is equivalent to reconstructing the cross section  as a function of ${\Delta}_\perp$, by detecting the recoiled proton, versus obtaining the cross section from the dijet only as a function of ${q}_\perp$.

To apply Eq.~(\ref{base}), we compute the leading order cross section $d\sigma_0$ from the Color Glass Condensate effective theory \cite{McLerran:1993ni,McLerran:1993ka,McLerran:1994vd,Gelis:2010nm,Albacete:2014fwa}, including energy evolution by solving the leading order Jalilian-Marian-Iancu-McLerran-Weigert-Leonidov-Kovner JIMWLK equations numerically \cite{JalilianMarian:1996xn,JalilianMarian:1997gr,Ferreiro:2001qy,Iancu:2000hn}. This part is the same as that computed in Ref.~\cite{Mantysaari:2018zdd,Mantysaari:2019csc}, and more details can be found there.

\begin{figure}[tb]
\includegraphics[width=0.45\textwidth]{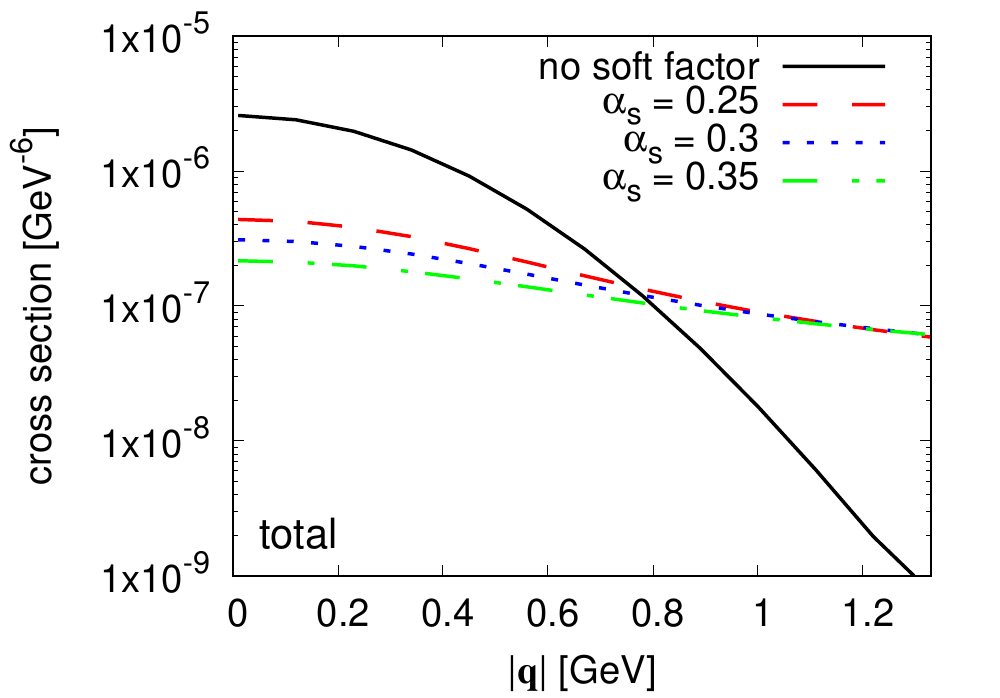}
\caption{Angle integrated total ($\sigma_T+\sigma_L$) dijet cross section from the CGC for $W=140$  GeV, $Q^2=25$ GeV${}^2$, $|{P}|=3.5$ GeV, $z=\bar{z }=0.5$. We show
 results, neglecting soft final state radiation and resummation of NGL's (black solid lines), contrasted with results where soft radiation and NGL's are included for fixed coupling $\alpha_s=0.25,0.3,0.35$.}\label{fig:EIC:CS}
\end{figure}

A typical EIC kinematic is applied: $E_p=250~\rm GeV$ for the proton beam energy, $W=\sqrt{(P+q)^2}=140$ GeV for the center of mass energy and $ Q^2=25~\rm GeV^2$ for the photon virtuality. %Our results are presented in the analysis frame, where photon and proton have no transverse momentum, while the nucleon is right-moving with a large $P^+$component. 
We consider symmetric u/d-flavor dijets with $z=\bar{z}=0.5$, where $z=k_1^-/q^-$ is the longitudinal momentum fraction of the first jet relative to the photon and $\bar{z}=1-z$. In \Fig{fig:EIC:CS}, we %show dijet production cross sections for transversely (a) and longitudinally (b) polarized photons, as well as the total cross section (c) as a function of $|{q}_\perp|$ for fixed dijet momentum $|{P}|$,
%integrated over the azimuthal angles of the dijets. We 
compare results of the total cross section ($\sigma_T+\sigma_L$) from the CGC computation without final state radiation (for which $|{q}_\perp|=|{\Delta}_\perp|$) (black solid lines) with computations including soft radiation. % (red dashed , blue dotted and green dashed-dotted lines). 
The latter is obtained by employing \Eq{base} with the soft factor $S(\lambda_\perp)$ given by \Eq{eq:softfactordef}. We compare with three different coupling constant: $\alpha_s=0.25$ (red dashed lines), $0.3$ (blue dotted lines) and $0.35$ (green dashed-dotted lines), respectively. 

The effect of soft radiation on the cross section is similar to that in \Fig{fig:E791}. The un-convoluted cross section %is of the order of $2\cdot 10^{-6}$ ($10^{-8}$) GeV ${}^2$ for transverse (longitudinal) photons and 
falls steeply at larger $|\mathbf{q}_\perp|$. Soft final state radiation reduces the cross sections by roughly a factor $5-10$ in the back-to-back limit at small $|{q}_\perp|$. As the back-to-back peak is smeared
by the soft radiation, the cross section at large $|\mathbf{q}_\perp| \gtrapprox 0.8$ GeV is larger than the un-convoluted one. To produce the results of \Fig{fig:E791} a coupling of $\alpha_s=0.3$ was assumed. Here, we vary the coupling between $\alpha_s =0.25-0.35$ to provide a systematic uncertainty of our results.

\begin{figure}[tb]
\includegraphics[width=0.45\textwidth]{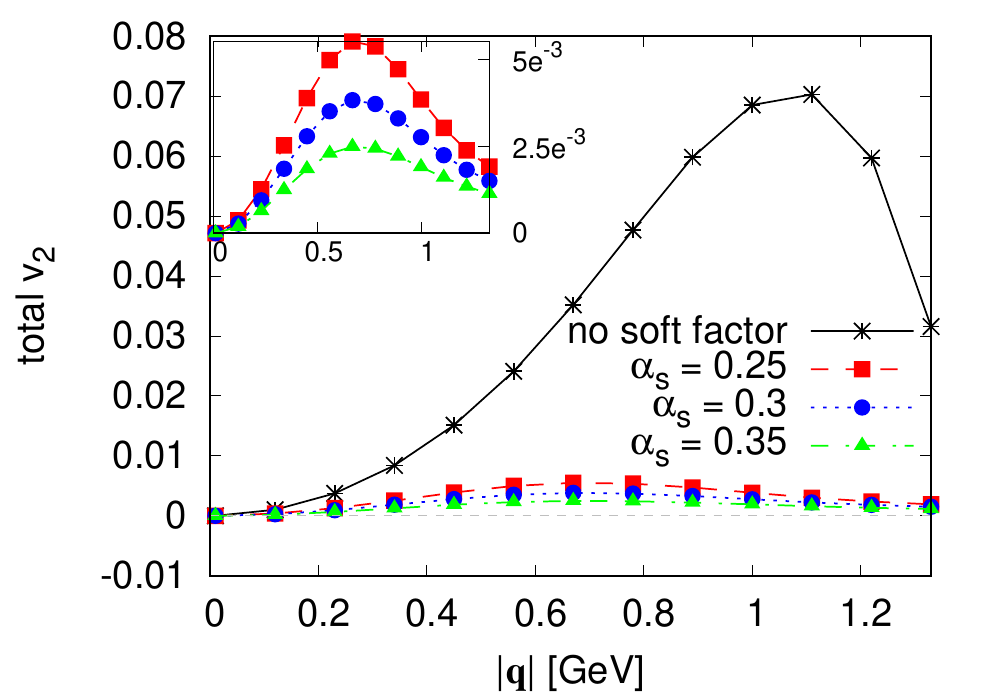}
\caption{Elliptic Fourier coefficients $v_2$ of the dijet cross section for $W=140$  GeV, $Q^2=25$ GeV${}^2$, $|{P}_\perp|=3.5$ GeV, $z=\bar{z }=0.5$, plotted as a function of $|{q}_\perp|$. }\label{fig:EIC:v2}
\end{figure}

More importantly, the soft factor effects are different for the two contribution terms in the differential cross section of Eq.~(\ref{eq:csazimuthal}). Therefore, there will be net effects on the azimuthal modulation of $\text{d}\sigma / \text{d} |{q_\perp}|$. 
In \Fig{fig:EIC:v2}, we show the azimuthal modulation $v_2$ of the total cross section ($\sigma_T+\sigma_L$)
 as a function of $|{q}_\perp|$.
In the presented kinematical regime, %where $W=140$  GeV, $Q^2=25$ GeV${}^2$, $|{P}|=3.5$ GeV, $z=\bar{z }=0.5$, 
the $v_2$ obtained from the un-convoluted cross section, e.g. from $\text{d}\sigma / \text{d} |{\Delta}_\perp|$ (black stars), shows a strong modulation of up to %$v_{2} \approx 4 \%$ for transverse and  
$v_{2} \approx 7\%$. % for longitudinal photons. %The total $v_2$ is dominated by the transverse component and is at most $v_2 \approx 7\%$ at around $|{q}_\perp|=|{\Delta}_\perp|\approx 1$ GeV. 
In contrast, including soft gluon radiation significantly reduces the resulting azimuthal modulation of $\text{d}\sigma / \text{d} |{q_\perp}|$. Here, too we show results for different values of $\alpha_s = 0.25,0.3,0.35$. % (red circles, blue squares and green triangles ), consistent with the analysis of the E791 data presented in the previous section. 
At $\alpha_s=0.3$ the maximal modulation is %$v_{2,T}\lessapprox 0.5\%$ for transverse and  
$v_{2,L}\lessapprox 0.5\%$. % for longitudinal photons.%, while the total is $v_2 \approx 0.5 \%$ at $|{q}_\perp|\approx 0.7$ GeV. 
Similar trends are observed when studying the transverse and longitudinal contributions separately.
The rather strong suppression of the total $v_2$ due to soft radiation will make it very difficult to measure for experiment. This means that the dijet total momentum cannot be a proxy for the recoiling proton momentum. Measuring the recoiled target directly is imperative to extract information about parton Wigner distributions at the EIC.

%\nm{I have the data for $W=900$  GeV, $E_p=2500$ GeV, $Q^2=60$ GeV${}^2$, $|P_\perp|=5$ GeV on tape and have some nice plots. Appendix or next chapter?}

%\section{Summary and Discussions}
{\it Summary and Discussions. }
%\nm{need to write story here}
In this paper, we %have investigated the soft gluon radiation associated with the final state jets in the hard diffractive dijet production processes. All order resummation of the soft factor was derived following recent theoretical developments on NGL contributions in jet processes.
% We 
have demonstrated that the soft factor from all order resummation plays an important role to describe the total transverse momentum distribution for the $\pi$-induced diffractive dijet production from E791 collaboration. This provides an important test of the resummation formalism. 

Similar effects have been found for the diffractive dijet production in $e+p$ collisions at the future EIC. Especially, the azimuthal angular modulation of $v_2$ is strongly suppressed with soft factor contribution. Therefore, we need to measure the recoil nucleon momentum to observe the sizable $v_2$ for the diffractive dijet production and from that we can extract the Elliptic gluon distribution, a non-trivial gluon tomography at small-$x$. 

In addition, we emphasize that the comparison between the measurements in terms of $\vec{q}_\perp$ and $\vec{\Delta}_\perp$ (as shown in Figs.~\ref{fig:EIC:CS} and \ref{fig:EIC:v2}) provides a unique opportunity to study the QCD resummation effects, which can be compared to other jet production processes~\cite{Dasgupta:2001sh,Dasgupta:2002bw,Banfi:2002hw,Banfi:2003jj,Hatta:2009nd,Hatta:2013iba,Neill:2018mmj}. Finally, we point out that the extension to $e+A$ collisions at the EIC should be done accordingly, where we also have to take into account the coherent and incoherent diffractive contributions. We plan to address this in a separate publication.

\section*{Acknowledgements}
%\acknowledgements
We thank R.~Boussarie, F.~Salazar, and B.~Schenke for discussions. F.Y. thanks V.~Braun and G.~Miller for discussions on the E791 experiment. 
This material is based upon work supported by the LDRD programs of 
Lawrence Berkeley National Laboratory and Brookhaven National Laboratory, the U.S. Department of Energy, 
Office of Science, Office of Nuclear Physics, under contract numbers 
DE-AC02-05CH11231 and  DE-SC0012704. N.M. is funded by the Deutsche Forschungsgemeinschaft (DFG, German Research Foundation) - Project 404640738.  This research used resources of the National Energy Research Scientific Computing Center (NERSC), a U.S. Department of Energy Office of Science User Facility operated under Contract No. DE-AC02-05CH11231.

\end{document}